\documentclass[11pt]{article}
\usepackage{moriond,epsfig}
\usepackage{url}

\def\Journal#1#2#3#4{{#1} {\bf #2}, #3 (#4)}

\def\PRL{\em Phys. Rev. Lett.}
\def\PRD{{\em Phys. Rev.} D}

\def\mnras{\em MNRAS}

\def\cqg{\em Classical and Quantum Gravity}
\def\aaps{\em Astronomy \& Astrophysics Supplement}
\def\astrophys{\em Astrophys. J}
\def\pasp{\em PASP}
\def\pasa{\em PASA}
\def\RPPh{\em Reports on Progress in Physics}

\def\be{\begin{equation}}
\def\ee{\end{equation}}
\def\bea{\begin{eqnarray}}
\def\eea{\end{eqnarray}}

\begin{document}
\vspace*{4cm}
\title{SEARCHING FOR ELECTROMAGNETIC COUNTERPARTS OF GRAVITATIONAL WAVE TRANSIENTS}
       
\author{M. BRANCHESI}
\address{on behalf of the LIGO Scientific Collaboration and the Virgo Collaboration\\
DiSBeF, Universit\`a degli Studi di Urbino ``Carlo Bo'', 61029 Urbino, Italy\\
INFN, Sezione di Firenze, 50019 Sesto Fiorentino, Italy}
\author{A. KLOTZ}
\address{Universit\'e de Toulouse,
Institut de Recherche en Astrophysique et Plan\'etologie\\
9 Avenue Colonel Roche,
31028 Toulouse Cedex 4
France}
\author{M. LAAS-BOUREZ}
\address{ICRAR / School of Physics, University of Western Australia,\\ 
Crawley WA 6009, Australia}

\maketitle

\abstracts{ A pioneering electromagnetic (EM) observation follow-up
  program of candidate gravitational wave (GW) triggers has been
  performed, Dec 17 2009 to Jan 8 2010 and Sep 4 to Oct 20 2010,
  during the recent LIGO/Virgo run.  The follow-up program involved
  ground-based and space EM facilities observing the sky at optical,
  X-ray and radio wavelengths. The joint GW/EM observation study
  requires the development of specific image analysis procedures able
  to discriminate the possible EM counterpart of GW trigger from
  background events. The paper shows an overview of the EM follow-up
  program and the developing image analysis procedures as they are
  applied to data collected with TAROT and Zadko.}
\section{Introduction}

The LIGO~\cite{Abb} and Virgo~\cite{Ac} detectors aim at the first
direct detection of gravitational waves from very energetic
astrophysical events. The most promising sources are mergers of
neutron stars (NS) and/or stellar mass black holes (BH) and the core
collapse of massive stars. More exotic sources include cosmic string
cusps.  It is likely that a fraction of the large energy reservoir
associated to those sources be converted into electromagnetic
radiation. This possibility is a feature of several astrophysical
scenarios. For instance, Gamma-Ray Bursts (GRBs) are thought to be
associated with the coalescence of NS-NS or NS-BH binaries or the
collapse of very massive stars (see~\cite{Aba1} and references
therein). Another scenario associated with compact object mergers is
the prediction~\cite{Me} of an isotropic EM emission from
supernova-like transients powered by the radioactive decay of heavy
elements produced in merger ejecta (this is referred to as the
\textit{kilonova} model). There are models that predict that cusps
produce electromagnetic radiation~\cite{Va}.

In this respect, multi-messenger GW and EM astronomy is a very
promising field of research. An electromagnetic counterpart discovered
through a follow-up of a gravitational wave candidate event would
considerably increase the confidence in the astrophysical origin of
the event. The detection of an EM counterpart would give the precise
localization and possibly lead to the identification of the host
galaxy and redshift. Furthermore, EM and GW observations provide
complementary insights into the progenitor and environment physics. In
the long term combined measurements of the luminosity distance through
GW radiation and redshifts through EM observations may allow a new way
of estimating some cosmological parameters.

\section{Enabling EM Follow-up of Candidate GW Events}

\subsection{Selection of Candidate GW Events}

A first program of EM follow-up to GW candidates took place (Dec 17
2009 to Jan 8 2010 and Sep 4 to Oct 20 2010) during the last
LIGO/Virgo observation periods, thanks to the development of a
low-latency GW data analysis pipeline that uses real time
gravitational wave triggers to obtain prompt EM observations to search
for the EM counterparts.

One of the challenges of successfully obtaining ``target of
opportunity'' EM observations is to identify the GW candidates
quickly: the data from the three operating detectors (the two LIGOs
and Virgo) must be transferred and analyzed in near-real time.  As
soon as the data become available, three search algorithms (Omega
Pipeline, coherent Wave Burst both described in~\cite{Ab2,Se} and
Multi Band Template Analysis~\cite{Bu}) run over the data.  For each
generated trigger, the direction of arrival of the wave (and hence
potential sky position of the source) is estimated using a method
based on differences in arrival time at each detector.  The event
candidates are collected in the Gravitational-wave candidate event
database (GraCEDb). Two software packages LUMIN and GEM select
statistically significant triggers and determine the telescope
pointing positions. This process typically takes $\sim$10 minutes. It
is followed by a manual event validation. A team of trained experts is
on duty and their role is to rapidly coordinate with scientists at the
GW detectors to evaluate the detector performances. If no problem is
found the alert is sent to telescopes.  The entire process is
typically completed within 30 minutes.  The triggers selected as GW
candidates for EM follow-up are the ones detected in triple
coincidences and with a power above a threshold estimated from the
distribution of background events. A full description of the GW
trigger selection and the entire EM follow-up process is detailed
in~\cite{Em}.
\subsection{Sky Pointing strategy}

The uncertainty in the source direction reconstruction scales
inversely with the signal-to-noise ratio~\cite{Fa}. GW events near the
detection threshold are localized into regions of tens of square
degrees. Generally, the error regions have a non-trivial geometrical
shape, often formed of several disconnected patches.  Follow-up
EM-telescopes with a wide Field Of View (FOV) are thus required.
However, the majority of those telescopes have a FOV which is much
smaller than the GW angular error box.
Additional priors are necessary to improve the location accuracy and
increase the chance that the actual source be in the selected FOV.
The observable Universe is limited to an horizon of 50 Mpc, taking
into account the detector sensitivity to the signals coming from NS
binaries~\cite{Ab3}.  The observation of the whole GW error box is not
required~\cite{Ka}, but it can be restricted to the regions occupied
by Globular Clusters and Galaxies within 50 Mpc, listed in the
Gravitational Wave Galaxy Catalog~\cite{Wh}. Tens of thousands of
galaxies are included within this horizon and the GW observable
sources are more likely to be extragalactic.

To determine the telescope pointing position, the probability sky map
based on GW data is ``weighted'' taking into account the mass and the
distance of nearby galaxies~\cite{Nu} and globular clusters. It is
assumed that the probability of a given galaxy being the host of the
actual source i) is directly proportional to the galaxy's mass (the
blue luminosity is used as proxy for the mass and thus for the number
of stars) and, ii) is inversely proportional to the distance.

\subsection{Follow-up EM Observatories and Observation Strategy}

The follow-up program involved ground-based and space EM facilities:
the {\it Liverpool Telescope}, the {\it Palomar Transient Factory}
(PTF), {\it Pi of the Sky}, {\it QUEST}, {\it ROTSE III}, {\it
SkyMapper}, {\it TAROT} and the {\it Zadko Telescope} observing the
sky in the optical band, the {\it Swift} satellite with X-ray and
UV/Optical telescopes and the radio interferometer {\it LOFAR}. The
observing strategy employed by each telescope will be described
in~\cite{Em}.

The cadence of EM observations is guided by the expected EM
counterpart. The optical afterglow of an on-axis GRB peaks few minutes
after the EM/GW prompt emission.  The kilonova model predicts an
optical ligth curve that peaks a day after the GW event, due to the
time that the out flowing material takes to become optically thin.
The agreement with the EM facilities allowed observations as soon as
possible, the day after the GW event and, repeated observations over
longer time-lag to follow the transient light curve dimming.

During the recent winter and summer LIGO/Virgo runs a total of 14
alerts have been sent out to the telescopes and 9 of them led to 
images being taken.

\section{Optical Transient Search in the Wide-Field Telescope Observations}
Once the follow-up observations are completed, the collected set of
images needs to be analyzed to decide the presence or not of an
optical transient of interest. The analysis method is conceptually
similar to one used to study a GRB afterglow with a main difference:
the arc minute localization of the current generation gamma-ray
observatories allows a significant reduction of the search area with
respect to the GW observations.

Searching for optical transients in a large sky area requires the
development and use of specific image analysis procedures able to
discriminate the EM counterpart from background/contaminant events.
Several analysis pipelines are being developed and tested by groups
within the LIGO Scientific Collaboration and the Virgo Collaboration
in partnership with astronomers.

This section describes one of the considered approaches based on the
cross-correlation of object catalogs obtained from each image.  The
resulting pipeline has been designed and tested with the images
collected by the two {\it TAROT} and the {\it Zadko} telescopes.

{\it TAROT}~\cite{Kl} are two robotic 25 cm telescopes with a FOV of
3.5 square degrees located in Calern (France) and in La Silla
(Chile). In case of a GW alert, {\it TAROT} followed a nominal
observation schedule including six consecutive images with 180 second
exposures during the first night and same for the three following
nights. An exposure of 180 seconds corresponds to a red limiting
magnitude of 17.5 under ideal conditions.  {\it Zadko}~\cite{Co} is a
1 meter telescope with a FOV of 0.17 square degrees located in Gingin
(Western Australia). For each GW trigger, {\it Zadko} followed a
nominal observation schedule including a mosaic of five fields with
six consecutive images during the first night and same for the three
following nights.  A 120 sec exposure corresponds to a red limiting
magnitude of 20.5 under ideal conditions.

The main steps of the fully automated analysis pipeline are as follows:\\
\noindent
1) extraction of the catalog of objects visible in the images using
  \texttt{SExtractor}~\cite{Be};\\
\noindent
2) removal of ``known objects'' listed in USNO-A2.0 or USNO-B star
  catalogs that are complete down to a fainter magnitude than the
  collected images by using a positional cross-correlation tool
  \texttt{match}~\cite{Ri};\\
\noindent
3) trace objects in common to several image catalogs by using a
  cross-positional check. This results in a light curve for each traced
  object;\\
\noindent
4) rejection of ``rapid contaminating transients'' (like cosmic rays,
  asteroids or noise): the presence is required in at least four
  consecutive images;\\
\noindent
5) rejection of ``background transients'': the objects are selected in
  the image regions associated with the galaxies within 50 Mpc.  A
  circular region with a diameter equal to 4 times the major axis galaxy
  size is used.  This is to take into account the possible offset
  between the host galaxy center and the optical transients (observed up
  to tens of kpc for GRBs);\\
\noindent
6) rejection of ``contaminating events'' like galaxies, variable stars
  or false transients by analyzing the light curves. The code selects
  the objects that show a luminosity dimming with time. Assuming that
  the dimming is described by a single power-law ${\cal L} \propto
  t^{-\beta}$, corresponding to a linear variation in terms of magnitude
  equal to $m = 2.5 \beta \log_{10}(t) + C$, a ``slope index''
  $2.5\beta$ is defined and evaluated for each objects. The expected
  ``slope index'' for GRB afterglows and kilonova-like light curves is
  around 2.5-3. In practice, a conservative cut is applied by selecting
  as the possible EM counterparts the objects with the ``slope index''
  larger than 0.5. This value has been checked using Monte Carlo
  simulations.

The preliminary results on the pipeline sensitivity indicate for a
survey red limiting magnitude of 15.5 that the majority of GRB
afterglows can be detected further away the GW horizon distance of 50
Mpc, while the kilonova objects can be detected up to a distance of 15
Mpc.  These results are obtained by repeatedly running the pipeline
over sets of {\it TAROT} and {\it Zadko} images where fake on-axis GRB
and kilonova optical transients were injected.

Estimates of the rate of false detections may be deduced from the
occurence of detected optical transients that would be unrelated to
the GW event and observed by chance in the field. The contaminant
transients that are able to pass the light-curve cut include some
rapid variable Cepheid stars and Active Galactic Nuclei.
\vspace{-0.5em}
\section{Concluding remarks}
\vspace{-0.5em}
The present paper reports on the first EM follow-up program to GW
candidates performed by the LIGO/Virgo collaborations together with
partner observatories.  This follow-up program is a milestone toward
the advanced detector era. With a ten-fold improvement of
sensitivity~\cite{Ab4}, the number of detectable sources increases by
a factor of $10^3$. It is likely that advanced detectors will make the
first direct detection of GWs. The observation of an EM counterpart
may be a crucial ingredient in deciding the astrophysical nature of
the first event.

\end{document}